# Electrospun Conjugated Polymer/Fullerene Hybrid Fibers: Photoactive Blends, Conductivity through Tunnelling-AFM, Light-Scattering, and Perspective for Their Use in Bulk-Heterojunction Organic Solar Cells

*Zhenhua Yang,*[†,1] *Maria Moffa*[||,1] *Ying Liu,*[†] *Hongfei Li,*[†] *Luana Persano,*[||] *Andrea Camposeo,*[||] *Rosalba Saija,*[#] *Maria Antonia Iatì,*[§] *Onofrio M. Maragò,*[§] *Dario Pisignano,*[*,||,‡,2] *Chang-Yong Nam,*[¶] *Eyal Zussman*[††] *and Miriam Rafailovich*[*,†]

[†]*Department of Materials Science and Engineering, State University of New York at Stony Brook, Stony Brook, New York, 11794-2275 (USA).*

[||]*NEST, Istituto Nanoscienze-CNR, Piazza S. Silvestro 12, I-56127 Pisa (Italy)*

[#]*Dipartimento di Scienze Matematiche e Informatiche, Scienze Fisiche e Scienze della Terra, Università di Messina, viale F. Stagno D'Alcontres 31, I-98166 Messina (Italy)*

[§]*CNR-IPCF, Istituto per i Processi Chimico-Fisici, viale F. Stagno D'Alcontres 37, I-98166 Messina (Italy)*

[‡]*Dipartimento di Matematica e Fisica "Ennio De Giorgi", Università del Salento, via Arnesano, I-73100 Lecce (Italy).*

[¶]*Center for Functional Nanomaterials, Brookhaven National Laboratory, Upton, New York, 11973-5000 (USA)*

[††]*Department of Mechanical Engineering, Technion-Israel Institute of Technology, Haifa 32000 (Israel)*

[1]Z.Y. and M.M. contributed equally to this work.

[2]Present Address: Dipartimento di Fisica, Università di Pisa and CNR Istituto Nanoscienze, Largo B. Pontecorvo 3, I-56127 Pisa (Italy)

* Corresponding authors: Prof. Dario Pisignano, e-mail: dario.pisignano@unipi.it. Prof. Miriam Rafailovich, e-mail: miriam.rafailovich@stonybrook.edu

**ABSTRACT**

Hybrid conjugated polymer/fullerene filaments based on MEH-PPV/PVP/PCBM are prepared by electrospinning, and their properties assessed by scanning electron, atomic and lateral force, tunnelling, and confocal microscopy, as well as by attenuated total reflection Fourier transform-infrared spectroscopy, photoluminescence quantum yield and spatially-resolved fluorescence. Highlighted features include ribbon-shape of the realized fibers, and the persistence of a network serving as a template for heterogeneous active layers in solar cell devices. A set of favorable characteristics is evidenced in this way in terms of homogeneous charge transport behavior and formation of effective interfaces for diffusion and dissociation of photogenerated excitons. The interaction of the organic filaments with light, exhibiting specific light-scattering properties of the nanofibrous mat, might also contribute to spreading incident radiation across the active layers, thus potentially enhancing photovoltaic performance. This method might be applied to other electron donor-electron acceptor material systems for the fabrication of solar cell devices enhanced by nanofibrillar morphologies embedding conjugated polymers and fullerene compounds.

**Introduction**

Bulk heterojunction (BHJ) polymer solar cells are a field of intense interest due to their flexibility, low cost, and ease of processing.[1-4] The performance of these devices is known to be highly dependent on the morphology of the active layers,[5-9] and to critically suffer from unfavorable features such as isolated domains or dead ends.[4] Therefore a variety of methods have been explored to finely control such morphology, such as thermal annealing,[8,10-12] solvent annealing,[13,14] self-assembly of columnar polymer phases,[15] and micropatterning.[16] Recently, a different route has been developed based on the use of organic nanofibers to optimize the donor-acceptor interface morphology. Polymer nanorods[12] and nanofibers,[8] prepared by melt-assisted wetting[12] or by a variety of thermally-assisted processing techniques,[8,17-19] have been shown to increase hole mobility, inducing higher mesoscopic order and crystallinity, enhanced donor-acceptor interfaces, and ultimately leading to better device performance compared with polymer films, due to the formed nanoscale and possibly interpenetrating network of fibrillar components. Systems investigated so far have been mostly limited to melts or self-assembled nanostructures made of polythiophene donors, such as core-shell nanorods or layers embedding fibrils of poly(3-hexylthiophene) or poly(3-butylthiophene) and the acceptor phenyl-$C_{61}$-butyric acid methyl ester (PCBM).[8,12,18-20] Developing processing methods which can extend the range of used conjugated polymers and offer higher throughput are therefore strongly desirable, in view of exploiting networked fibrillar morphologies to enhance the performance of BHJ solar cells at larger scale.

Electrospinning is a straightforward technology for the realization of continuous fibers with sub-μm diameter through the application of a high voltage bias to polymer solutions.[21,22] This process enables the production of non-wovens made of fibers with high surface-area-to-volume ratio, which can be deposited in random networks or in uniaxially aligned arrays with three-dimensional porosity. Due to its versatility in terms of usable polymers and blends and to its good throughput, electrospinning shows high potential as method to tailor the microstructure and the composition of active materials for polymer optoelectronic devices, including solar cells, in a controlled way.[23-25] Demonstrated applications include field-effect transistors[26,27] and luministors,[28] rechargeable batteries,[29] energy harvesters,[30,31] light-emitting devices[32-34] and nanopatterned lasers.[35] Electrospun photovoltaic materials would ultimately lead to the development of so-called solar cloths[36] for smart textile technologies. In addition, organic fibers with diameters matching the wavelength of solar light can exhibit exceptional light-scattering performances, as recently found in insect scales formed by interconnected filaments of chitin,[37] thus possibly enhancing the coupling of light into the absorbing regions of layered devices. Unfortunately, electrospinning conjugated polymers is frequently difficult due to the generally poor viscoelastic behavior of the solutions, which corresponds to a low amount of molecular entanglements, and to the limited solubility of these compounds. Hence the use of electrospinning methods to produce fibrillar components in BHJ polymer solar cells and modules has been only rarely attempted.[38-42] Recently it was shown that electrospinning can be carried out with a few photoactive polymers, upon encasing them in an inert/insulating polymer in a core-shell fiber configuration.[39,41] This method though also proved to be difficult since the shell polymer must first be removed for the device to function, and this step could potentially damage the surface of the remaining functional fibers. Much wider efforts have been instead directed to the realization of dye-sensitized solar cells exploiting electrospun materials. A recent review can be found in Ref. 43.

Here we focus on polymer blends to facilitate the electrospinning process. We have previously demonstrated that addition of a secondary polymer component is beneficial to maintain the structure of BHJ solar cell films, despite the fact that the second component is not photoactive.[15] In this work we report on electrospun poly[2-methoxy-5-(2-ethylhexyloxy)-1,4 phenylenevinylene]/ polyvinylpyrrolidone/PCBM (MEH-PPV/PVP/PCBM) blend fibers which are stable, easily produced, and yet photoactive, and which significantly improve the performance of BHJ solar cells. Fibers are incorporated in the device active layer upon deposition of a fully solubilized poly(3-hexylthiophene-2,5-diyl)/PCBM (P3HT/PCBM) backfill layer[39] on top. Partial redissolution of the electrospun fibers occurs at the backfill layer deposition, however the structural coherence and shape of the fibers are largely maintained in solution-processed multilayer device architectures.[39] The optical and conduction properties of the three-dimensional entangled fiber networks are studied by complementary spectroscopies as well as lateral-force and tunneling atomic force microscopy (AFM), highlighting the formation of compound interfaces driving exciton dissociation in the organic filaments, together with homogeneous charge transport and generation of streaks along the longitudinal axis of fibers, primed by PCBM inclusions. The light-scattering of these networks is also investigated to evidence their capability to direct photons across the mat plane, thus potentially enhancing photo-absorption from the UV to the infrared. The network of organic fibers, serving as a template for P3HT/PCBM deposition, would be also promising in view of the incorporation of the hybrid, electrospun conjugated
polymer/fullerene in organic photovoltaic devices.

## Experimental Section

**Materials.** MEH-PPV (Mw=150-250 kDa) is purchased from Aldrich. PCBM is provided by
SES research and PVP (Mw=1,300 kDa) by Alfa Aesar. Chlorobenzene, chloroform and P3HT
(Mw 54-75 kDa) are obtained from Sigma-Aldrich. All materials are used without further
purification.

**Fabrication and characterization of the photoactive blend nanofibers.** The solution for
electrospinning is made of MEH-PPV/PVP/PCBM dissolved in chloroform at a concentration of
15/15/10 mg/mL, respectively. It is then stirred at room temperature for 12 hours to allow for
complete polymer dissolution, loaded in a syringe with a 21 gauge stainless steel needle, and
injected through the needle at constant flow rate (0.5 mL/h) by a syringe pump (Harvard
Apparatus, Holliston, MA). A 8 kV voltage is applied at the needle using a high-voltage power
supply (EL60R0.6–22, Glassman High Voltage, High Bridge, NJ). Electrospinning is carried out
in ambient atmosphere, and fibers are collected on square (1.5×1.5 cm$^2$) polymer/TiO$_2$/indium
tin oxide (ITO)/glass substrates mounted on a rotating disk collector (4000 rpm) at a distance of
15 cm from the needle. A cross-bar pattern of fibers is obtained by depositing fibers along two
mutually perpendicular directions, rotating the substrates by 90° after the first deposition stage,
as inspected by scanning electron microscopy (SEM, FEI, Hillsboro). Prior to electrospinning,
the substrates are carefully prepared as described in the next Section.

UV-Visible absorption spectra are obtained with a Thermo Scientific Evolution 200 UV-VIS
Spectrophotometer. The photoluminescence (PL) properties of the fibers are examined in micro-

PL mode by using a confocal microscopy system, composed of an inverted microscope (Eclipse Ti, Nikon) and a laser scanning head (A1R MP, Nikon). To this aim, samples are excited by an Ar+ laser ($\lambda_{exc}$=488 nm) through a 20× objective (numerical aperture = 0.5), while the intensity of the fluorescence, collected by the same excitation objective, is measured by using a spectral detection unit equipped with a multi-anode photomultiplier (Nikon). This allows spatially-resolved spectra and fluorescence images to be collected.

PL quantum yield measurements are performed following the procedure reported in Ref. 44. Fiber samples deposited on quartz substrates were positioned in an integrating sphere and excited by a UV light emitting diode (peak emission wavelength = 300 nm and linewidth = 18 nm). The excitation and emission optical signals are collected by an optical fiber, coupled to a monochromator (iHR320, Jobin Yvon) and measured by a charge coupled device camera (Symphony, Jobin Yvon). Attenuated Total Reflectance Fourier Transform-Infrared (FTIR) spectroscopy is performed on electrospun fibers using a spectrometer (Spectrum 100, Perkin Elmer, Waltham, MA) equipped with a ZnSe crystal for coupling (Perkin Elmer). Images of the surface topography and electrical conductivity are also obtained using a Bruker Dimension Icon AFM (Multimode, Bruker) operating in PeakForce tunneling mode.

**Devices.** ITO-coated glass is polished in UV/ozone for 10 min to remove any organic impurity. A $TiO_2$ solution is prepared according to previous reports.[45] Briefly, the solution is obtained by dissolving 1mL $Ti(OC_4H_9)_4$ in 10 mL ethanol, followed by adding 1 mL of $CH_3COOH$, then 1 mL of acetylacetone, and 1 mL of deionized water. The solution is stirred at room temperature for 30 min before each reagent is added. A 30 nm thick $TiO_2$ layer is then spun onto ITO at 3000 rpm for 20 s and baked in air at 400°C for 2 hours on a hot plate. To maximize adhesion of fibers, a ~10 nm thick film of MEH-PPV/PVP/PCBM is spin-cast on the $TiO_2$ layer at 6000 rpm for 30 s. MEH-PPV/PVP/PCBM fibers are then electrospun as described above. A solution of P3HT/PCBM (15/9 mg/mL in chlorobenzene) is used to interconnect the fibers upon spin-coating at 1000 rpm for 30 s. Samples are then annealed at 150 °C for 10 min in a vacuum oven. The resulting thickness of the active layer is (121±7) nm. Finally, the devices are completed by thermal evaporation of 8 nm $MoO_3$ and 100 nm Ag electrode with a Kurt J. Lesker PVD 75 vacuum deposition system at Brookhaven National Laboratory. Control devices are prepared with the same procedure but without electrospun fibers, with thickness (112±5) nm. The performance of so-realized solar cells is tested by a 150 W solar simulator (Oriel) with an AM 1.5G filter for solar illumination. The light intensity is adjusted at 100 mW $cm^{-2}$ by a calibrated thermopile detector (Oriel).

**Light scattering at nanofibers.** Light scattering calculations for the developed material are carried out in the T-matrix formalism. We consider the fiber structure as composed of aggregates of spheres embedded into a homogeneous, isotropic, indefinite medium. The optical properties of the subunit spheres and the surrounding medium are calculated using the Bruggeman description[46] (see Supporting Information for details). The incident field is the polarized plane wave (whose results are eventually averaged over the in-plane polarization angle), hence the total field outside the particle is the sum of the incident and scattered field. The scattered field is obtained applying the boundary conditions across the surface of each particle of the structure, linking the internal and external fields. The scattering problem is solved by the T-matrix method,[47-52] based on the definition of a linear operator relating the incident field to the scattered field.[47] In brief, the starting point of the method is the field expansion in terms of the spherical multipole fields, i.e., the vector solutions of the Maxwell equations in a homogeneous medium that are simultaneous eigenfunctions of the angular momentum and the parity operators. The operator $S$, called transition operator, is introduced thanks to the linearity of the Maxwell equations and of the equations expressing the boundary conditions across the surface of the particle. The representation of the operator $S$ on the basis of the spherical multipole fields gives the T-matrix whose elements, $S_{lml'm'}^{(pp')}$, (with $p$ parity index, $l$ and $m$ angular momentum indices) contain all the information about the scattering process, but are independent of the state of polarization of the incident field. Here, we use the cluster model[48-50] to get the scattering properties of fibrous mats embedded in a surrounding medium. The cluster model is a special case since the T-matrix approach solves the scattering problem by an aggregate (cluster) of spheres without resorting to any approximation.[50] This is highly useful since it allows to simulate many situations of practical interest where one deals with non-spherical scatterers. We also highlights that in order to calculate the T-matrix of a cluster one has to solve a linear system of equations with, in principle, infinite order. Thus, the system must be truncated to some finite order by including into the multipole expansions terms up to this truncation, chosen so to ensure the convergence of the calculations. As a consequence, the computational demand for these calculations increases with the cube of the number of spheres.[51-53]

## Results and Discussion

In our samples, a backfill layer of fully solubilized P3HT/PCBM is spin-cast onto the top of the nanofibrous layer, thus interconnecting the electrospun fibers and avoiding short circuit contacts of sandwiching electrodes, which would be likely to occur due to the highly porous nature of the network of filaments. P3HT/PCBM as a backfill layer has very good film-forming properties, and through two different types of photoactive polymers (i.e. MEH-PPV and P3HT) the resulting devices would be able to increase their acceptance range of incoming light frequencies. Finally, $MoO_3$ and Ag are deposited as electron blocking layer and anode, respectively. For the deposition of the nanofibrous layers, electrospinning is optimized to provide a three-dimensional and interconnected structure made of MEH-PPV fibers blended with PCBM. To overcome the poor spinnability of pristine MEH-PPV, due to its limited molecular weight (150-250 KDa) and generally poor viscoelastic behaviour in solution, an easily spinnable component given by PVP is added to the solution. MEH-PPV and PVP are blended at a 1:1 (w:w) relative concentration, which is found to lead to fiber formation with stable electrified jets and good efficiency. In addition, we tune the density of the nanofibrous network to obtain a given degree of coverage onto the surface. A deposition time of a few minutes is used to obtain a thickness ranging from one to a few superimposed layers of fibers. Differently from previous works,[39,41] we do not employ sacrificial sheaths in core-shell nanofibers, which would need additional processing steps to be removed, but instead use a unique blend system relying on percolative paths for internal charge transport. The morphology of MEH-PPV/PVP/PCBM electrospun fibers, shown in Fig. 1, highlights that three-dimensional nonwoven networks are formed, where fibers are uniformly dispersed and exhibit ribbon-shape, and transversal size ~300-450 nm, with very rare beads over a large area. The domains observed on the fiber surface suggest phase separation of the different polymer components at scales (few tens of nm) well matching the typical exciton diffusion lengths in organics.[4]

FTIR spectra of electrospun MEH-PPV/PVP/PCBM fibers show a band peaked at 528 cm$^{-1}$, which is characteristic for the fullerene derivatives[54,55] (Fig. S1 in the Supporting Information file). More importantly, the presence of PCBM in these fibers and the formation of effective internal interfaces between the acceptor and the donor compounds in the organic filaments is supported by the reduction of the PL quantum yield found for fibers with fullerene compared to MEH-PPV/PVP fibers. Indeed, the PL quantum yield of the MEH-PPV/PVP fibers, measured by an integrating sphere and accounting for the number of emitted photons per incident photons,[44] is 11±1%, decreasing by at least one order of magnitude (<1 %) upon fullerene addition. Furthermore, the PL spectrum of the MEH-PPV/PVP/PCBM fibers is blue-shifted by 10 nm, compared to MEH-PPV/PVP fibers (Fig. 2a). Such blue-shift is also found by measuring the spatially-resolved fluorescence spectra of the fibers by micro-PL, allowing us to collect PL maps with sub-micron spatial resolution (Fig. 2b,c). The uniform brightness along the fibers in the micrographs in Fig. 2b and 2c clearly indicates a homogeneous incorporation of MEH-PPV in the electrospun filaments. In particular, the analysis of the spectra collected from different areas of the fiber samples evidence a substantial blue-shift of the average peak wavelength (by 12 nm, Fig. 2d,e) due to the presence of PCBM, together with a broadening of the peak wavelength distribution, featuring an increase of the full width at half maximum from 2 nm to 7 nm upon PBCM addition. Following photo-excitation, the emission properties of conjugated polymers, which are multi-chromophore systems composed by many active sub-units, are determined by energy migration, funneling the excitation toward those chromophores that have lower characteristic energies and that emit light.[56] In pristine electrospun nanofibers made by conjugated polymers this process occurs on picosecond timescales, it can be tailored by the degree of aggregation of the conjugated polymer chains, and it determines the emission to occur from the more extended and conjugated sub-units.[57] The presence of PBCM and the formation of effective interfaces for dissociation of photogenerated excitons through electron transfer[58] lowers the PL quantum yield. In addition, the blue-shifted emission indicates a decreased aggregation of the conjugated polymer chains in composite samples.[57,59] Overall, these findings support the occurrence of diffusion and dissociation of photogenerated excitons at interfaces formed within electrospun nanofibers.

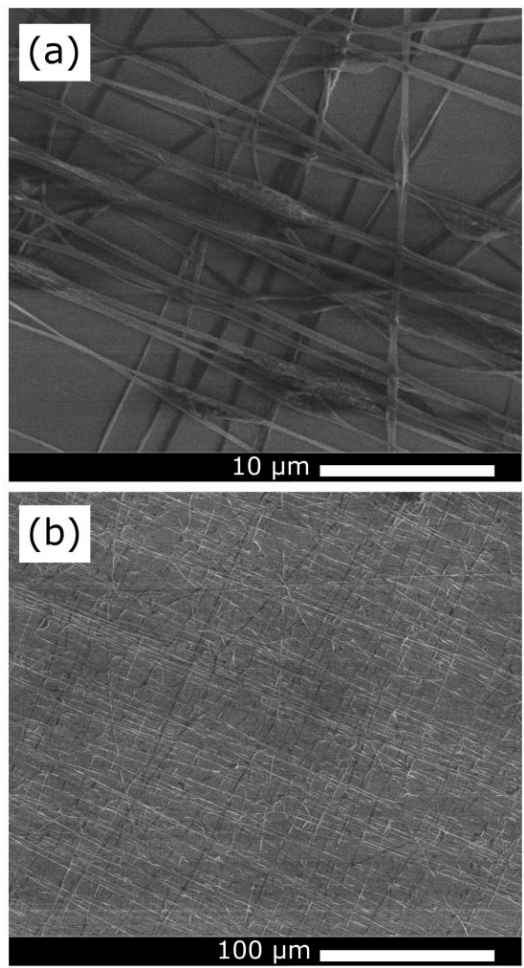

**Fig. 1.** SEM micrographs of electrospun MEH-PPV/PVP/PCBM fibers in cross-bar configuration, imaged at different magnifications. Locally rough surface, suggesting phase-separation occurring in the blend during electrospinning, is visible in the thicker regions in the fibers in (a). Images also show that fibers are superimposed at different heights on the substrate, thus generating a truly three-dimensional networks.

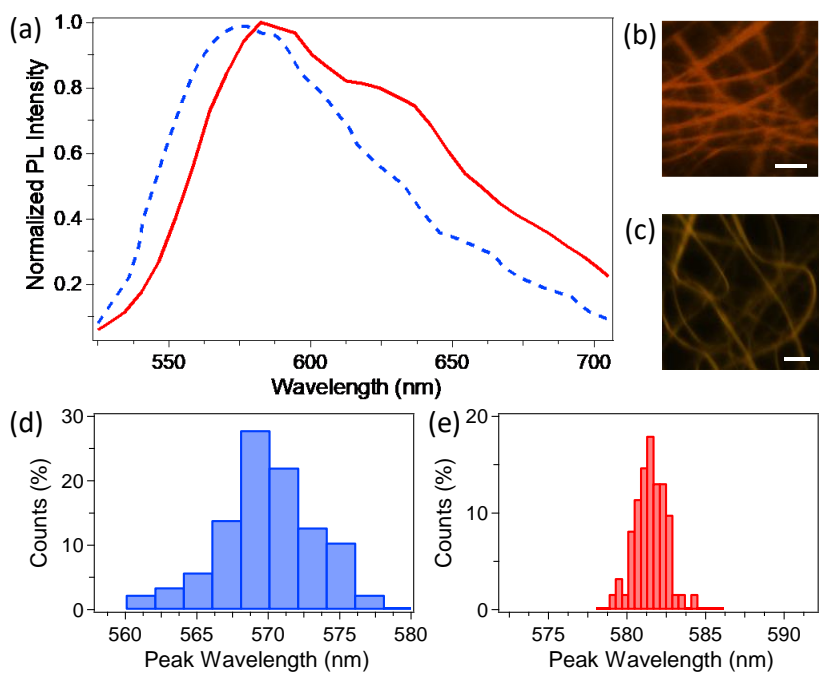

**Fig. 2.** (a) PL spectra of MEH-PPV/PVP nanofibers (red continuous line) and of MEH-PPV/PVP/PCBM nanofibers (blue dashed line). (b,c) Exemplary fluorescence maps of MEH-PPV/PVP and MEH-PPV/PVP/PCBM, respectively, measured by confocal microscopy. Scale bars: 10 μm. (d,e) Distribution of the peak emission wavelengths for MEH-PPV/PVP/PCBM and MEH-PPV/PVP nanofibers, respectively, obtained after measuring spatially-resolved fluorescence spectra. The spectra used for the analysis are averaged over a ~1 μm² area along the length of fibers.

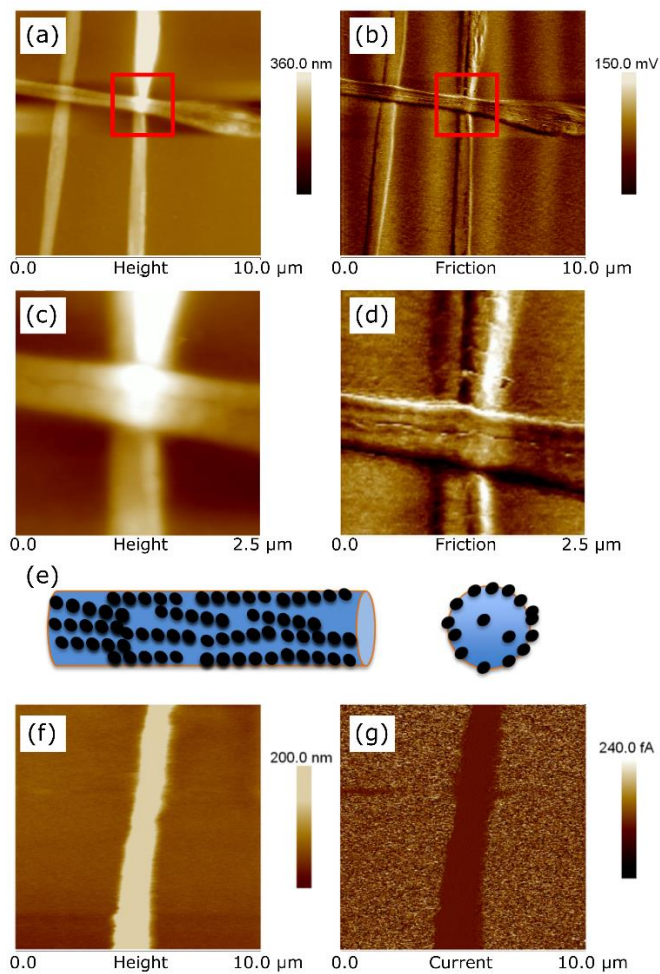

**Fig. 3.** AFM (a) and lateral force (b) micrographs (10 μm × 10 μm) of cross-bar MEH-PPV/PVP/PCBM fibers deposited on the blend film with addition of the backfill layer. (c) and (d) are magnified images for the region delineated by red squares in (a) and (b). (e) Schematic presentation of PCBM assembly at the fiber surface. The dark dots represent PCBM nanoparticles (features not to scale compared to the fiber diameter). (f, g) TUNA images (10 μm × 10 μm) of ITO-glass/TiO₂/ MEH-PPV/PVP/PCBM film/electrospun MEH-PPV/PVP/PCBM fibers samples. (f): Fiber topography and (g) corresponding current map with an applied voltage of 2 V.

The morphology of the active layer surface as resulting from the fiber deposition is studied with lateral-force AFM (Fig. 3). It can be seen from Fig. 3a that the structural coherence of MEH-PPV/PVP/PCBM fibers is well maintained after depositing the P3HT/PCBM backfill layer. Though with flattening due to partial dissolution, the fiber network still serves as a template[39] for the active layer (Fig. 3b). In this way a complex, double-heterojunction structure is kept in the active film, with tandem features due to the presence of diverse conjugated polymers increasing the range of effective absorption for incoming light. The layered structures create multiple donor-acceptor interfaces which are formed in the fibers and in the backfill layer, respectively, and a PCBM component possibly promoting the development of percolative paths for electrons across regions realized with different blends. In Fig. 3c and 3d we show a magnified view of the region highlighted by the squared in Fig. 3a and 3b, respectively. Though not visible in the topographic images due to their small heights (~2 nm), PCBM inclusions can be clearly seen in the friction image (Fig. 3d), where they appear slightly darker or harder than their surroundings. The PCBM is arranged in long streaks which run along the length of the fiber. The PCBM that we view also lies directly on the surface of the fiber, since the friction mode scans are only sensitive to differences in mechanical or adhesion surface forces. Imaging the electrospinning process using fast X-ray analysis, it was previously shown that particle inclusions self-assemble in long streaks that run along the length of fiber, and delineate the stream lines of the electrospining jet. In order to maintain lamellar flow in the viscous stream, particle inclusions are also pushed towards the surface of the stream, where the deformation they cause is minimized.[60] Along the jet, due to the dominant effect of axial stretching accompanied by lateral contraction, the particles tend to be segregated towards the surface of the fibers in the as spun samples. A cartoon of the PCBM distribution in the electrospun fibers, in both side and top view is shown in Fig. 3e. If the PCBM concentration is sufficiently high a percolative network forms on the surface of the fiber, which might be very effective in increasing the current of a BHJ photovoltaic device. We also utilize Tunnelling (TUNA)-AFM to investigate the fiber conductivity. Fig. 3f displays a region of an electrospun MEH-PPV/PVP/PCBM filament deposited on the blend film. The corresponding current map for zero bias highlights no significant difference between the fiber and the surrounding film (Fig. S2). Upon increasing the applied voltage to 2V, the fiber features are brought back and match well with the topography image (Fig. 3g). The current detected from the fiber area (about 1.5 fA) is lower than that from film due to the higher serial resistance. Indeed, the ratio of the measured current values for the fiber and the surrounding layer (~25) well agrees with the ratio of the corresponding thickness in the two probed areas. Overall, these results indicate a homogeneous charge transport behavior for the adhesion film and the fibers along their transversal direction, and conductivity values ($1.02$-$1.34 \times 10^{-7}$ S/m) adequate for optoelectronic applications, such as solar cells.

The transmission spectra of different samples (MEH-PPV/PVP/PCBM adhesion film, MEH-PPV/PVP/PCBM film with added backfill layer and MEH-PPV/PVP/PCBM film with fibers and backfill layer) are shown in Fig. 4. MEH-PPV/PVP/PCBM films exhibit broad absorption at 430-570 nm and maximum absorption at about 490 nm arising from the $\pi$-conjugated structure. In addition, the presence of PCBM can be appreciated due to the related absorption around 300 nm.[61]

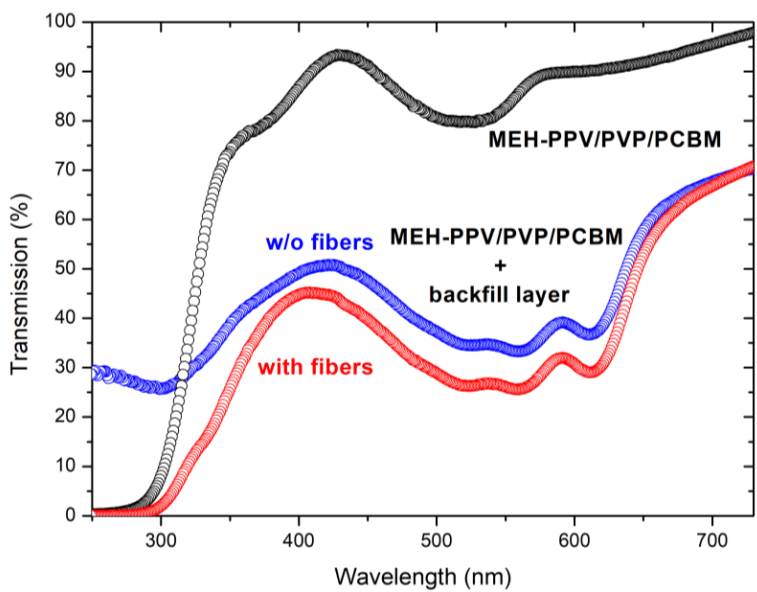

**Fig. 4.** Transmission spectra in the visible and near-infrared, for a MEH-PPV/PVP/PCBM thin film (black circles), a MEH-PPV/PVP/PCBM thin film with backfill layer (blue) and a MEH-PPV/PVP/PCBM thin film with nanofibers and backfill layer (red).

Following the addition of the backfill layer, extra peaks at 517 nm and 556 nm and one shoulder at 605 nm are appreciated, correlated to π–π* transitions from P3HT,[61] together with a significant decrease of the transmitted intensity. Research shows that there is no interaction between P3HT and MEH-PPV in terms of absorption spectra and the absorbed energy of the films of the blend series is independent of the blending ratio.[62] Therefore, the here found decrease of transmission through the layer is to be attributed to the incorporation of the nanofibers.

To elucidate in depth the underlying working mechanisms, the light scattering properties of the ordered fibers embedded in their external medium are described by exploiting the transition matrix (T-matrix) formalism.[47-49] The T-matrix approach combines an accurate description of the scattering process with computational efficiency and a wide particle size range when describing complex non-spherical and composite particles. Here, we model the composite polymer fibers as ordered clusters of spheres,[51,52,63] as shown in Fig. 5a. In order to mimic the non-spherical fiber section as well as the arrangement of the filaments used in devices we consider ordered arrays of sphere dimers with diameter of 300 nm, as the fiber short axis, and resulting transverse size of each dimer of 600 nm, corresponding to the transverse fiber size.

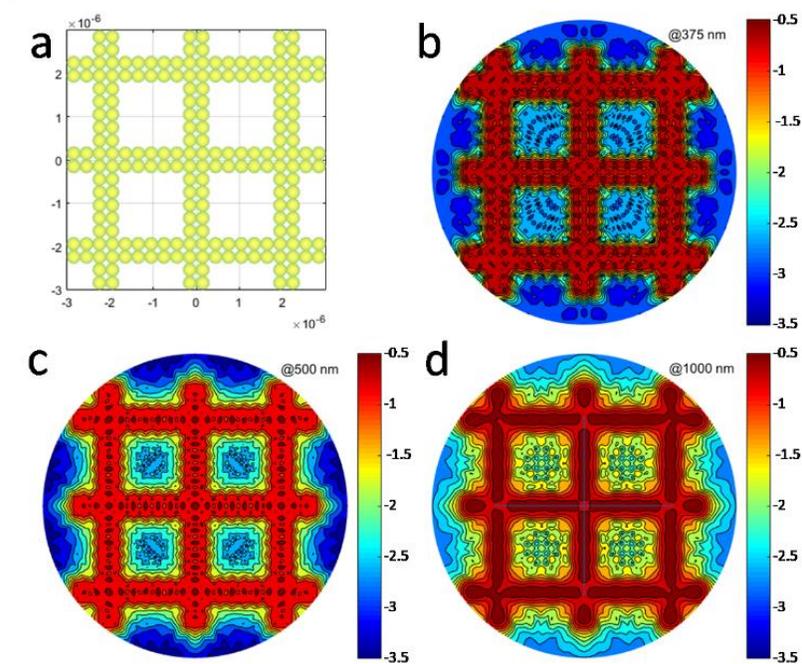

**Fig. 5.** (a) Sketch of the model structure for the light scattering calculations from the ordered, composite polymer nanofiber mats embedded in a dielectric medium. The organic filaments are modeled as aggregate of spheres with 300 nm diameter to match the thickness of the ribbon-shaped fibers realized in the experiments. The optical properties of the nanofibers and of the external medium are obtained using an effective medium theory (see text). (b-d) Normalized intensity maps ($|E_S/E_0|^2$, in logarithmic scale) of the scattered field for the fiber mats at different wavelentgths: 375 nm (b), 500 nm (c), and 1000 nm (d). In all the calculations the light propagates orthogonally to the fiber mats and with a polarization that is averaged in plane. The fiber structures scatter more strongly in the infrared because of the better-matching average thickness, with scattered light also being more spread out over the structure.

The optical constants of the fibers and of the surrounding medium are calculated using an effective medium theory exploiting the Bruggeman effective dielectric function for composite materials (Supporting Information),[47] where the different composition of the fiber and external medium is accurately taken into account. Light scattering maps for different wavelengths of the incident light, matching different regions of the solar spectrum (from the ultraviolet to the near infrared), are shown in Fig. 5b-d. For each wavelength, a map is obtained for the scattered field normalized to an unpolarized incident field intensity, $|E_S/E_0|^2$. From these maps it is clear that the fibrous structures scatter light in an efficient way, producing hot spots of radiation spreading across the involved photovoltaic interfaces. In addition, light scattering is stronger in the near infrared where the average thickness of deposited fibers is closer to the incident wavelength. Moreover, the scattered light at relatively longer wavelengths is spreads out over and outside the fibrous structure, thus reliably increasing optical coupling with the surrounding layer and ultimately photon absorption across the organics.

The fabrication process of a prototype BHJ solar cell embedding electrospun MEH-PPV/PVP/PCBM fibers is summarized in Fig. 6, with the corresponding current density-voltage curves shown in Fig. 7. With the inclusion of electrospun MEH-PPV/PVP/PCBM fibers into the active layer, the $J_{sc}$ and $FF$ values respectively increase by 0.39 mA cm$^{-2}$ (i.e., from 3.63 to 4.02 mA cm$^{-2}$) and 4.4% (from 41%) on average, which is indicative of enhanced photon absorption according to the above reported mechanisms and consequently enhanced free-charge generation.[64] No significant change is found for the open-circuit voltage ($V_{oc}$ =0.56-0.57 V), suggesting a negligible effect on shunt resistance.[42] Finally, the power conversion efficiency (PCE) is increased by about 20% upon nanofiber embedment (from ~0.8 to ≥1%) inset of Fig. 7). The performance improvement is attributed to the active layer template formed with electrospun nanofibers, which affects various properties of the overall device including the internal light-scattering properties as specified above, in turn enhancing internal absorption of incident photons. In addition, the nanofiber template not only serves as a guide path for electron transport but also increases the interfacial area between donor and acceptor to induce more exciton dissociations. Such template approach is promising in view of application to other polymer solar cell systems in order to achieve higher efficiency under ambient conditions.

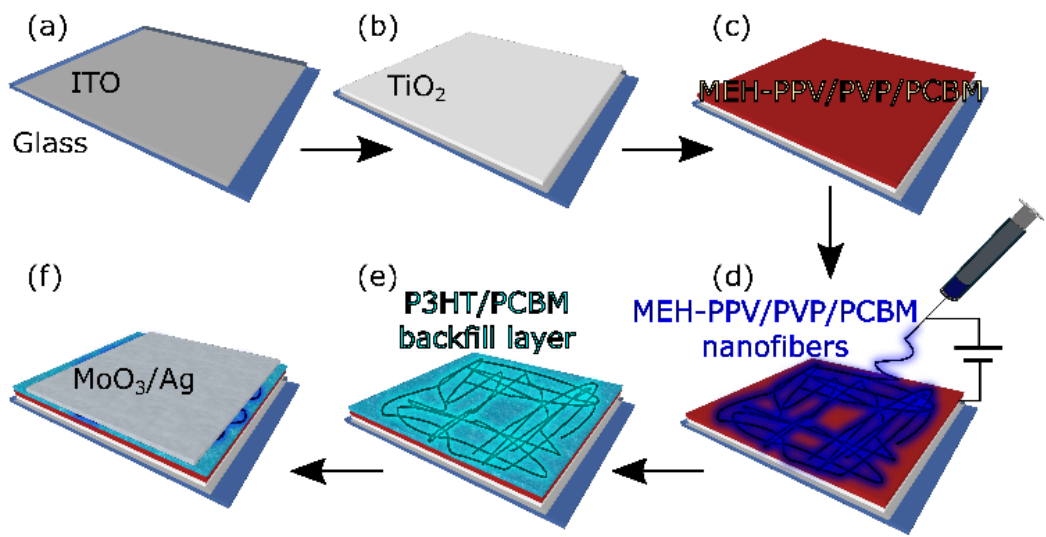

**Fig. 6.** Scheme of the process for realizing solar cell devices based on electrospun, MEH-PPV/PVP/PCBM nanofibers. The fibers are spun on ITO/glass substrates (a) following the deposition of a 30 nm thick $TiO_2$ layer (b) and of a spin-cast MEH-PPV/PVP/PCBM film (c). After electrospinning, (d), P3HT/PCBM is spin-cast to define the template active layer, and the electron-blocking and top electrode are thermally evaporated (f).

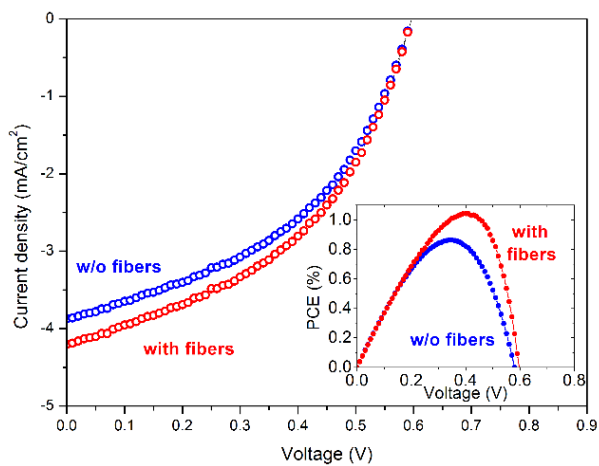

**Fig. 7.** Current density-voltage characteristics for devices with active layers of MEH-PPV/PVP/PCBM film without nanofibers (blue symbols) and with nanofibers (red symbols). Inset: corresponding PCE curves for exemplary devices without (blue symbols) and with nanofibers (red symbols).

**Conclusions**

In summary, MEH-PPV/PVP/PCBM nanofibers can be successfully prepared via electrospinning and incorporated into the active layer of polymer solar cell devices. The entangled fiber network serves as a template for the active layer, with effects promoting diffusion and dissociation of photogenerated excitons at involved organic interfaces as well as light-scattering redirecting incident photons across active layers. The electrospinning process forces particle inclusions towards the exterior of the spinning stream. As a result a particle-rich layer is formed with a well-defined pathway for electron conduction, as illustrated in Fig. 3e. This conformation could lead to increased current and fill factor upon including fiber structures in BHJ solar cells. So far observed performances from these architectures are largely in line with measurements reported for self-assembled polythiophene nanofibers in PCBM.[18] A body of literature exists on various optimized configurations exploiting blends of P3HT and PCBM, with PCE around 3% in average.[65] Here reported findings are relevant in this framework, since they can be easily generalized to other conjugated polymers, and since they highlight the synergy of phase-separation, transport, and light-scattering properties of organic nanostructures embedded in bulk heterojunction solar cells. Furthermore, various strategies can be figured out to further improve along these directions. Particularly, electrospun organic nanofibers can be easily doped or decorated by semiconducting quantum dots,[66,67] which has been found an excellent route to enhance donor-acceptor electronic interaction.[68] Also, electrospinning nanofibers based on conjugated polymers electrospun in controlled nitrogen atmosphere might lead to better charge-transport due to the reduced incorporation of oxygen during fabrication.[69] The methodology is likely to be extended to other donor-acceptor material systems, and to flexible solar cell devices as suggested by recent evidences on highly stable bendable field-effect transistors.[70]

## AUTHOR INFORMATION

**Corresponding Authors**

Dario Pisignano. E-mail address: dario.pisignano@unipi.it

Miriam Rafailovich, E-mail address: miriam.rafailovich@stonybrook.edu

## ACKNOWLEDGMENT

The research leading to these results has received funding from the European Research Council under the European Union's Seventh Framework Programme (FP/2007-2013)/ERC Grant Agreements n. 306357 (ERC Starting Grant "NANO-JETS"). M.R. and Z.Y. gratefully acknowledge the support of the National Science Foundation (NSF grant INSPIRE#1344267). The Apulia Networks of Public Research Laboratories Wafitech (09) and MITT (13), and the Advanced Energy Center-ThINC facility are also acknowledged.

# Electrospun Conjugated Polymer/Fullerene Hybrid Fibers: Photoactive Blends, Conductivity through Tunnelling-AFM, Light-Scattering, and Perspective for Their Use in Bulk-Heterojunction Organic Solar Cells

*Zhenhua Yang,[†,1] Maria Moffa[∥,1] Ying Liu,[†] Hongfei Li,[†] Luana Persano,[∥] Andrea Camposeo,[∥]*

*Rosalba Saija,[#] Maria Antonia Iatì,[§] Onofrio M. Maragò,[§] Dario Pisignano,[*,∥,‡,2] Chang-Yong*

*Nam,[¶] Eyal Zussman[††] and Miriam Rafailovich[*,†]*

## SUPPORTING INFORMATION

[†]*Department of Materials Science and Engineering, State University of New York at Stony Brook, Stony Brook, New York, 11794-2275 (USA).*

[∥]*NEST, Istituto Nanoscienze-CNR, Piazza S. Silvestro 12, I-56127 Pisa (Italy)*

[#]*Dipartimento di Scienze Matematiche e Informatiche, Scienze Fisiche e Scienze della Terra, Università di Messina, viale F. Stagno D'Alcontres 31, I-98166 Messina (Italy)*

[§]*CNR-IPCF, Istituto per i Processi Chimico-Fisici, viale F. Stagno D'Alcontres 37, I-98166 Messina (Italy)*

[‡]*Dipartimento di Matematica e Fisica "Ennio De Giorgi", Università del Salento, via Arnesano, I-73100 Lecce (Italy). E-mail:*

[¶]*Center for Functional Nanomaterials, Brookhaven National Laboratory, Upton, New York, 11973-5000 (USA)*

[††]*Department of Mechanical Engineering, Technion-Israel Institute of Technology, Haifa 32000 (Israel)*

[1]Z.Y. and M.M. contributed equally to this work.

[2]Present Address: Dipartimento di Fisica, Università di Pisa and CNR Istituto Nanoscienze, Largo B. Pontecorvo 3, I-56127 Pisa (Italy)

* Corresponding authors: Prof. Dario Pisignano, e-mail: dario.pisignano@unipi.it. Prof. Miriam Rafailovich, e-mail: miriam.rafailovich@stonybrook.edu

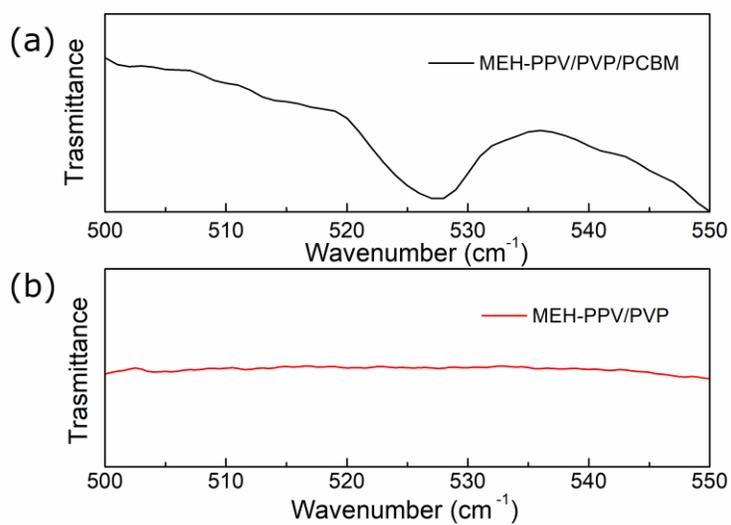

**Figure S1.** FTIR spectrum in the 500-550 cm$^{-1}$ region, for (a) MEH-PPV/PVP/PCBM fibers and (b) MEH-PPV/PVP fibers, highlighting the presence of the band peaked at 528 cm$^{-1}$ upon fullerene doping.

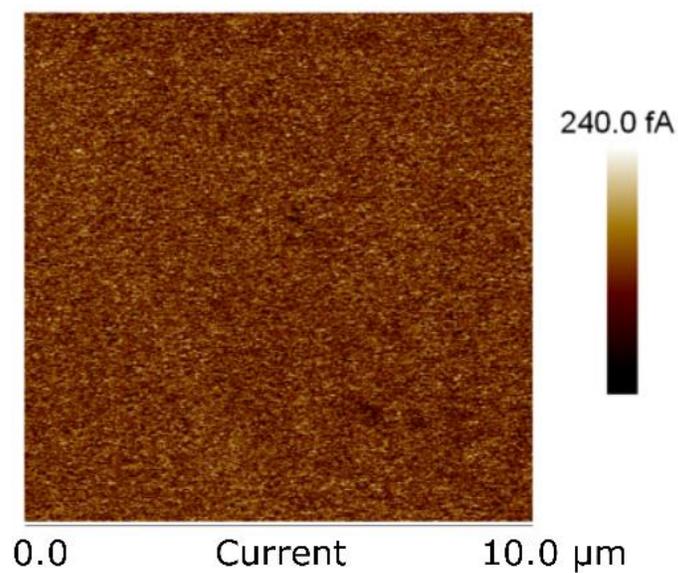

**Figure S2.** TUNA (10 μm × 10 μm) current map measured on a ITO-glass/TiO₂/MEH-PPV/PVP/PCBM film/electrospun MEH-PPV/PVP/PCBM fibers sample as that in Fig. 3f, with an applied voltage of 0 V.

**Effective medium theory and Bruggeman dielectric function.** In order to accurately calculate the optical properties of the different composite materials, nanofibers (MEH-PPV/PVP/PCBM) and surrounding medium (P3HT/PCBM), we exploit an effective medium theory approach and use the Bruggeman mixing rule.[S1] The Bruggeman effective dielectric function generally applies to a randomly inhomogeneous medium, that is a medium in which no distinguishable inclusions are present. Thus, the effective dielectric function, $\epsilon_{eff}$, for nanofibers or the surrounding medium is calculated separately starting from the dielectric function of their component materials as:

$$\sum_j V_j \frac{\epsilon_j - \epsilon_{eff}}{\epsilon_j + 2\epsilon_{eff}} = 0$$

where $V_j$ is the volume fraction occupied by the different components of dielectric constant $\epsilon_j$.